\documentclass[a4paper]{article}

\usepackage{INTERSPEECH2022}
\usepackage{multirow}

\title{Minimizing Sequential Confusion Error in Speech Command Recognition}
\name{Zhanheng Yang, Hang Lv, Xiong Wang, Ao Zhang, Lei Xie$^{*}$}
\address{
  Audio, Speech and Language Processing Group (ASLP@NPU),\\
School of Computer Science, Northwestern Polytechnical University, Xi'an, China
\thanks{Lei Xie is the corresponding author.}}
\email{\{zhhyang,zhangao1998\}@mail.nwpu.edu.cn,hanglv@nwpu-aslp.org,xwang@npu-aslp.org,\\lxie@nwpu.edu.cn}

\begin{document}
\maketitle
\begin{abstract}
  Speech command recognition (SCR) has been commonly used on resource constrained devices to achieve hands-free user experience. However, in real applications, confusion among commands with similar pronunciations often happens due to the limited capacity of small models deployed on edge devices, which drastically affects the user experience. In this paper, inspired by the advances of discriminative training in speech recognition, we propose a novel minimize sequential confusion error (MSCE) training criterion particularly for SCR, aiming to alleviate the command confusion problem. Specifically, we aim to improve the ability of discriminating the target command from other commands on the basis of MCE discriminative criteria. We define the likelihood of different commands through connectionist temporal classification (CTC). During training, we propose several strategies to use prior knowledge creating a confusing sequence set for similar-sounding command instead of creating the whole non-target command set, which can better save the training resources and effectively reduce command confusion errors. Specifically, we design and compare three different strategies for confusing set construction.  By using our proposed method, we can relatively reduce the False Reject Rate~(FRR) by 33.7\% at 0.01 False Alarm Rate~(FAR) and confusion errors by 18.28\% on our collected speech command set.
\end{abstract}

\vspace{-3pt}
\section{Introduction}
\vspace{-5pt}
Speech command recognition (SCR), which detects pre-defined commands from consecutive audio streams, is widely deployed on edge devices such as smart household appliances, control panels and various Internet of Things (IoT) gadgets. Similar to keyword spotting (KWS), an SCR system usually runs locally on the resource-limited edge device, and thus needs to be small-footprint and prompt. But differently and more challengingly, an SCR system often needs to detect various commands ranging from a dozen to over one hundred depending on the device and application. Thus, besides improving detection accuracy and suppressing false alarm, handling confusions on different commands (especially those with similar pronunciations) is also vital to the user experience.

HMM-based~\cite{1989Continuous, 1990A,2016Multi, 2017Compressed, 2018Monophone} approaches have dominated SCR for quite a long time, which usually use Gaussian Mixture Model (GMM) or Deep Neural Network (DNN) for acoustic modeling, and a Viterbi decoder is employed to search on the WFST decoding graph encoded by the command set to output the most likely command.
% to map acoustic features to the modeling units such as phonemes, word-pieces, syllables, etc ~\cite{2016Multi, 2017Compressed, 2018Monophone} at first. Then, a Viterbi decoder is employed to search on the WFST-based decoding graph, which encodes the target commands, lexicon, context decision tree and HMM topology information together, to output the most likely command. 
Recently, some researchers also proposed to use a so-called end-to-end (E2E) neural model for SCR, which discards the decoding process in the HMM approach and directly maps the acoustic features to sub-word or word/command units for detection. The E2E approaches, such as those adopt convolution neural network (CNN) based models ~\cite{0Convolutional}, usually have low computational consumption and simple system building procedure~\cite{0Convolutional,2014Small,wei2021end}. However, with the growth of command vocabulary,  
%but the number of target commands is often limited to a small set. This is because with the growth of command vocabulary, 
there is more variability in the length of commands. It is troublesome in context modeling for commands with different length, as fixed context (e.g. reception field in CNN) is often adopted. Apparently, it is also not flexible for new command expansion for customization purpose.  

%difficult to process contextual utterances and custom commands because of the large granularity of the modeling units.

In practical applications, SCR systems have to face the confusion errors between the target command and the other competitive commands due to (partially) similar pronunciation among commands, for example, ``turn on the air conditioner'' and ``turn off the air conditioner'' will result in totally different actions. However, the common training criteria, such as cross-entropy (CE) and connectionist temporal classification (CTC)~\cite{2006Connectionist, yan2020crnn}, usually focus on optimizing the posterior probability of target frame or sequence only, which do not optimize the discrimination among different commands. 
% Further more, SCR on edge devices ask for low latency and small-footprint model, which lead to particularly serious confusion errors.
% The confusion errors is particularly serious for SCR on edge devices due to the low latency and small-footprint requirements. 

% However, The common training criteria, such as cross-entropy (CE) and connectionist temporal classification (CTC)~\cite{2006Connectionist, yan2020crnn}, used in HMM-based or E2E models usually focus on optimizing the posterior probability of target frame or sequence only and do not explicitly take confusion error reduction as an optimization goal, leading to serious confusion errors.

% discriminative training criterion has been frequently used in speech recognition, which not only optimizes the accuracy of the target sequence but also suppresses the competitive non-target sequences. Common discriminative training criteria such as Maximum Mutual Information (MMI)~\cite{2016Purely,Chen_2018}, Minimum Phone Error (MPE)~\cite{2002Minimum,0Investigating}, Minimum Word Error Rate (MWER)~\cite{2016Minimum,guo2020efficient}, Minimum Classification Error (MCE)~\cite{1992Discriminative,4032780} have been widely adopted, leading to improved speech recognition performance. 

In this paper, we particularly address the command confusion problem by adopting discriminative training in SCR. Specifically, we propose a Minimize Sequential Confusion Error (MSCE) criterion for SCR, which is inspired by the MCE criterion and aims to improve the accuracy of command recognition and reduce confusion errors at the same time. The highlights of our proposed method are two-fold:
\begin{itemize}
\vspace{-2pt}
\item {We introduce the CTC loss as a \textit{measure function}, which could better define the likelihood of each command and thus improve the discriminative ability of the target command from all the other commands.}
\vspace{-2pt}
\item{We design and compare three strategies for confusing set constructing, namely \textit{pronunciation similarity strategy} (PSS), \textit{random selection strategy} (RSS) and \textit{hybrid strategy} (HS), according to the prior knowledge for each command to suppress. These strategies can reduce the training resource consumption notably.}
\end{itemize}\par
\vspace{-2pt}
Our experiments show that, by using our proposed MSCE training criteria, we can relatively reduce FRR by 33.7\% at 0.01 FAR and confusion errors by 18.28\% on our collected evaluation set for SCR.

\vspace{-7pt}
\section{Related work}
\vspace{-5pt}
Sequence discriminative training criterion has been frequently used in speech recognition, which not only optimizes the accuracy of the target sequence but also suppresses the competitive non-target sequences at the same time. One of the most frequently-used criteria is Maximum Mutual Information (MMI)~\cite{2016Purely,Chen_2018}. The MMI criterion aims to enlarge the discrimination between the target sequence and the overall sequence set. The MMI family includes LF-MMI~\cite{2016Purely}, Minimum Word Error Rate (MWER)~\cite{2016Minimum,guo2020efficient} and Minimum Phone Error (MPE)~\cite{2002Minimum,0Investigating}. The MWER and MPE criteria introduce WER and PER in the optimization objective respectively as weight coefficient of the target sequence, which aims to suppress the WER/PER directly during training. Another frequently-used criterion is MCE~\cite{4032780}, which optimizes the classification error directly instead of the probability distribution.

% Daniel Povey proposed LF-MMI in ~\cite{2016Purely}, which have been widely adopted in ASR and leading to improved speech recognition performance. MMI criterion aim to improve the discrimination between target sequence and competitive sequence instead of only optimize the framewise accuracy. Another frequently-used method is Minimum Word Error Rate (MWER)~\cite{2016Minimum,guo2020efficient} or Minimum Phone Error (MPE)~\cite{2002Minimum,0Investigating}, which introduce WER/PER in the optimization objective. These discriminative training criterions could get promotion because WER/PER is the performance evaluation metric of ASR task and these criterions optimize WER/PER of the target hypothesized sequence as well as suppress WER/PER of other hypothesized sequences directly during training.

In the case of SCR, which can be considered as a typical classification task, we aim to improve the classification accuracy and minimize the command confusion. In this paper, we particularly modify the MCE criterion for speech command recognition with the purpose of command confusion minimization.
%Thus MWER/MPER criterion used in ASR task may not be the optimal choice for SCR task. In this work, we try to modify the MCE criterion to adapted SCR and use it as our optimization objective to reduce command confusion problem.

\vspace{-7pt}
\section{Model Architecture}
\begin{figure}[!h]
	\vspace{-10pt}
	\centering 
	\includegraphics[width=1.0\linewidth]{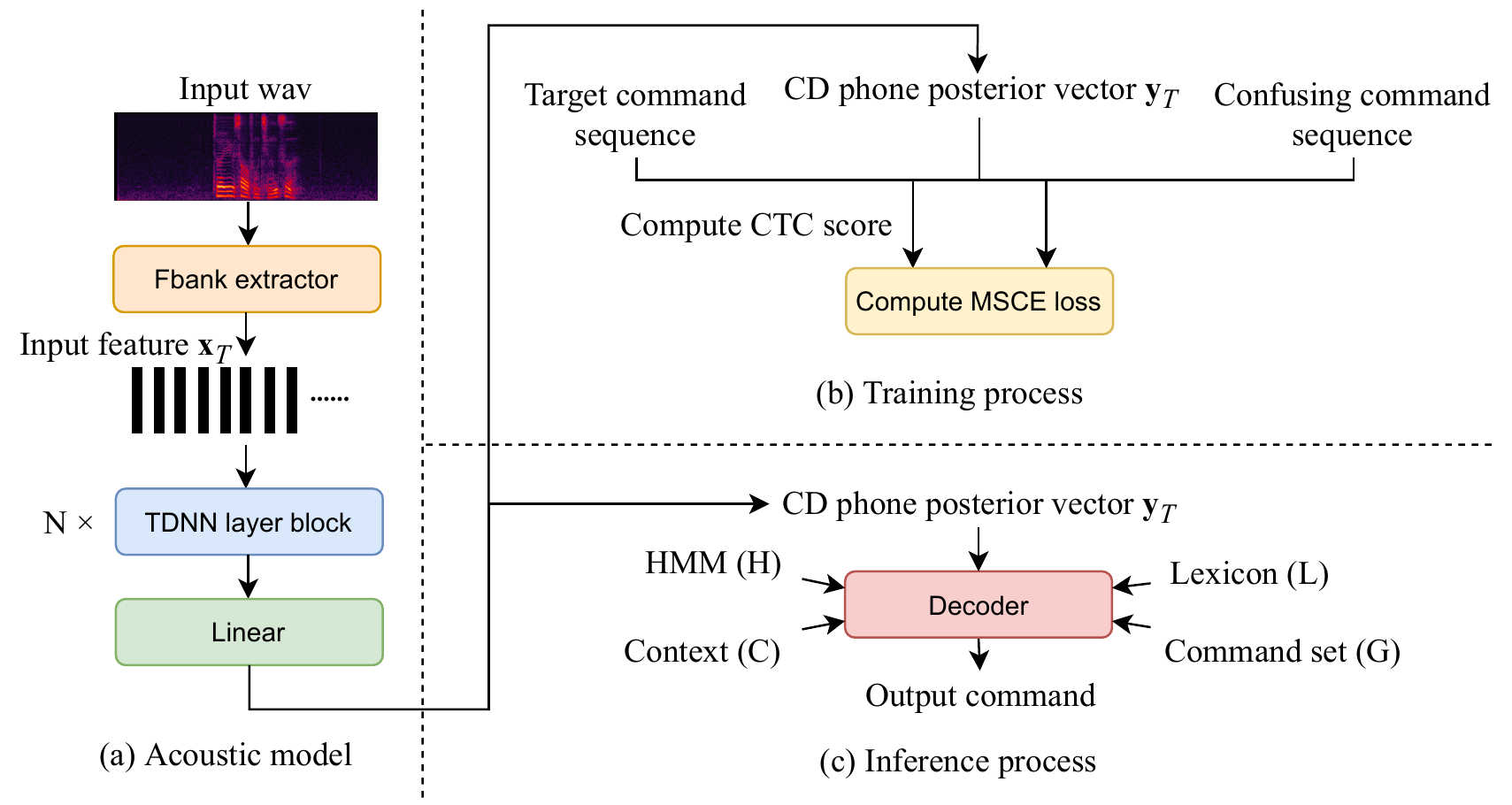}\vspace{-10pt}
	\caption{Overview on training and inference of our SCR system.}\vspace{-20pt}
	\label{fig1}
\end{figure}

\vspace{-5pt}
In this section, we introduce the overall framework of our system, as shown in Fig.~\ref{fig1}. Specifically, we deploy a context imbalanced time-delay neural network (TDNN), design the modeling units and simplify the decoding process.

\subsection{Acoustic Model}
\vspace{-0.5em}
In general, our acoustic model is established by multiple dilation TDNN-layer blocks as shown in Fig.~\ref{fig1}(a). Specifically, each block is stacked by a dilation TDNN layer, a batch normalization layer and a dropout layer.
We further improve the structure in the following two aspects: 1) To balance the receptive field and the model size, we deploy multi-scale (e.g. dilation rate=1,2,4) TDNN layers; 2) To keep low latency while maintaining good performance, we tune a few blocks in the middle of our model to only consider the left context.
In this way, we achieve a low latency and small footprint model which is easy to deploy on edge devices.

\vspace{-0.5em}
\subsection{Modeling Units and Decoding}
\vspace{-0.5em}
Considering the scalability of our model (e.g. custom commands) and the compatibility of varied lengths of commands, we use context-dependence~(CD) phones as the modeling units. The number of distinct states within the HMM for each CD-phone is predefined as 5. Notably, since the phoneme coverage of command recognition task is limited, we use a decision tree to bound the similar HMM states together, which is beneficial to restrict the model size on edge devices and improve the scalability of the model.

We compose various knowledge sources including target commands information~(G), lexicon~(L), context decision tree~(C) and HMM topologies~(H) together for decoding. The final HCLG decoding graph could be presented as
\vspace{-0.5em}
\begin{equation}
	\text{HCLG} = \min(det(H \circ C \circ L \circ G))~,
\end{equation}
%\vspace{-0.5em}
where $\min$ and $det$ represent minimization and determinization operations of WFST, respectively. %After composing and optimizing, different commands with similar pronunciation would share partial paths to cut down the size of decoding graph. 
The input label of the decoding graph is the bounded HMM state id and the output label is the word id in our setup.

During decoding, the aim of SCR is to figure out the most likely command in one-pass decoding. So our decoder performs Viterbi decoding through token passing algorithm on the search graph. Beam pruning is applied in practice. Compared to the existing token passing algorithm, some modifications are made in order to further reduce the consumption of memory and speedup:

\begin{itemize}
\vspace{-5pt}
\item{For SCR, we can obtain the length of each command in advance. So we use a fixed length array to record the output labels produced by decoding, instead of keeping a backward pointer for each token. In this way, we can get the recognized command directly from the array when triggered and avoid the procedure of generating lattice arcs and backtracing the whole decoding path. It helps to speedup and save memory.}
\vspace{-2pt}
\item {Without generating lattice, we just maintain two token lists dynamically for the current and next time steps. It helps to reduce the consumption of memory and time without sacrificing the performance.}
\end{itemize}
\vspace{-2pt}
Finally, once one of the decoding paths reaches the final state of the graph, the average cost on each state through this path is computed to compare with the threshold. If it is greater than the threshold, the system will trigger. The overall inference process, including decoding process, is shown in Fig~\ref{fig1}(b). In the next section we will introduce our discriminative training approach to suppress command confusion based on this model architecture.

\vspace{-5pt}
\section{Minimize Sequential Confusion Error}
\vspace{-5pt}
In this section, we introduce our proposed MSCE criterion inspired by the MCE criterion. We will first give a quick review on the MCE criterion. Then we introduce the CTC loss as our sequence-level measure approach, and further define our proposed MSCE training criterion.

%, and the discriminative training criterion, MCE, together to improve the discriminative ability of our acoustic model for dealing with confusing commands.
\vspace{-1em}
\subsection{Minimum Classification Error}
\vspace{-0.5em}
Generally speaking, the MCE criterion~\cite{4032780} is defined in terms of discriminative functions for each category is optimized. Compared with the maximum likelihood estimation (MLE) method, the MCE method minimizes the classification error directly rather than learning the true data probability distributions. So it is more appropriate for enlarging the discrimination among different categories. We introduce a three-step procedure to derive the MCE criterion as follows.

First, the form of the discriminative function is prescribed. For an input utterance $\textbf{x}_T$ with $T$ frames belonging to target category $\kappa$, we define the discriminative function for the target classification as $g_\kappa(\textbf{x}_T, \mathbf{\Lambda})$, in which $\mathbf{\Lambda}$ denotes the entire set of system parameters.
And we also deploy the same discriminative function for each non-target category, $\tau$, in non-target set, $S_{\tau}$, as $g_\tau(\textbf{x}_T,\mathbf{\Lambda})~(\tau \neq \kappa)$.

The second step is to define misclassification measure which allows us to embed the decision process into the overall minimum classification error formulation. The misclassification measure aims to distinguish the target category from all the other non-target categories for each input utterance $\textbf{x}_T$. A common used misclassification measure can be described as
\vspace{-0.5em}
\begin{equation}
	d_\kappa(\textbf{x}_T,\mathbf{\Lambda})=-g_\kappa(\textbf{x}_T,\mathbf{\Lambda})+ \sum_{\tau \in S_{\tau}}{g_\tau(\textbf{x}_T,\mathbf{\Lambda})},
%	\vspace{-1em}
\end{equation}
or
%\vspace{-1em}
\begin{equation}
	d_\kappa(\textbf{x}_T,\mathbf{\Lambda})=-\frac{g_\kappa(\textbf{x}_T,\mathbf{\Lambda})}{\sum_{\tau \in S_{\tau}}{g_\tau(\textbf{x}_T,\mathbf{\Lambda})}}~.
\end{equation}
The larger the $d_\kappa$ value, the better the discriminative performance.

Finally, we define a continuously differentiable loss function for the misclassification measure to optimize the parameters. It helps to eliminate the inconsistency between the scalar misclassification measure and the desired minimum error probability objective. We choose the most typical sigmoid function which is smoothed zero-one function suitable for gradient-based optimization algorithms:%, although it also can be substituted by exponential function and some piece-wise linear functions
\vspace{-0.5em}
\begin{equation}
	L(d_\kappa)=\frac{1}{1 + e^{-\xi(d_\kappa+\alpha)}}\quad ,\quad \xi>0~,
\end{equation}
where $\xi$ and $\alpha$ are constants to control the steepness and translation of the function.

\vspace{-1em}
\subsection{MSCE for Speech Command Recognition}
\vspace{-0.5em}
Inspired by the MCE discriminative criterion described above, we propose a Minimize Sequential Confusion Error (MSCE) criterion to deal with the command confusion problem in our SCR task. We define the \textit{measure function} in advance, and then embed it into the \textit{misclassification measure}. The model training process under MSCE criterion is shown in Fig~\ref{fig1}(b).

First, we regard the CTC loss function~\cite{2006Connectionist} as the sequence-level measure function. The CTC loss could be described as
\vspace{-0.5em}
\begin{equation}
	L_{CTC}(l|\textbf{x}_T,\mathbf{\Lambda})
	=\sum_{\pi \in B^{-1}(l) \atop \pi_{t} = l_s}{p(\pi|\textbf{x}_{t}, \mathbf{\Lambda})}
	=\sum_{s=1}^{\left|l^{'}\right|}{\frac{\alpha_t(s)\beta_t(s)}{y^t_{l^{'}_s}}}~,
\end{equation}
where $\pi$ corresponds to all reasonable alignment paths of the target label sequence, $l_s$ denotes the element in $l$ and $t$ denotes the time step. 
So given the target label sequence, the CTC criterion maximizes the sum of likelihood of all the possible alignment paths of the target sequence by the forward-backward algorithm. Besides, the phone-level sequence of corresponding command is regarded as the CTC label sequence in our setup.

%Besides, CTC function is smooth over the entire range and suitable for gradient algorithms, so we do not need to define additional loss functions. We define the misclassification measure as
After defining the sequence-level measure function for commands, we design the misclassification measure which aims to minimize the confusion errors. Equivalently, the measure aims to maximize the discrimination between the target command, $\kappa$, and all confusing commands in the confusion set, $\psi \in S_{\psi}$. So we define the misclassification measure for confusion errors in speech command recognition as
%\vspace{-1em}
\begin{equation}
	d_\kappa(\textbf{x}_T,\mathbf{\Lambda})=\frac{g_\kappa}{\sum_{\psi \in S_{\psi}}{g_\psi}}~,
\end{equation}
where the numerator part of the measure represents the sequence-level likelihood of the target command, while the denominator part describes the sequence-level likelihoods of the confusing commands. In this way, the better we minimize our sequence-level misclassification measure, the better the discrimination among commands.

After embedding the sequence-level measure function, CTC, into the misclassification measure, we achieve the proposed MSCE criterion as:
\vspace{-1em}
\begin{equation}
	d_\kappa(\textbf{x}_T,\mathbf{\Lambda})=\frac{L_{CTC}(\kappa)}{\sum_{\psi \in S_{\psi}}L_{CTC}(\psi)}~.
\end{equation}
Notably, our proposed MSCE criterion is a continuously differentiable function which can be optimized by gradient-decent conveniently.

\vspace{-1em}
\subsection{Confusing Command Selection}
\vspace{-0.5em}
In the training of MSCE criterion, it is not practical to include all non-target commands into the corresponding confusing set because CTC forward-backward algorithm needs to take up a large amount of GPU memory. Alternatively, we have to pick a finite number of non-target confusing commands to construct the corresponding confusing set. Here we propose three strategies.
\begin{itemize}
\item \textbf{Pronunciation similarity strategy (PSS):} Since command confusion errors usually happen among commands which have similar pronunciation, we choose confusing commands according to the phone similarity. Specifically, we expand all commands to phone-level according to the lexicon at first. Then, for each target command, we compute its  phone-level Levenshtein distance with each non-target command. At last, the top-$N$ most similar commands are selected to build the corresponding confusing set. Clearly, the confusing set of each target command is pre-defined before training.
\item \textbf{Random selection strategy (RSS):} Different from PSS, we randomly choose $N$ commands to construct the corresponding confusing set for each target command. Notably, the confusing set of each target command is generated dynamically during the training process.
\item \textbf{Hybrid strategy (HS):} This strategy combines both the characteristics of the two strategies described above. To be specific, we prepare two pools, one for top $N$ most similar commands with target command, the other for the whole commands set except for the target command. We randomly pick $i$~(0 $\leq$ $i$ $\leq$ $N$) commands from one pool and $(N-i)$ commands from the other to construct the corresponding confusing set dynamically during the training process.
\end{itemize}

\vspace{-1em}
\subsection{Multi-Task Learning}
\vspace{-0.5em}
%We deploy multi-task training for metric criteria except cross entropy baseline, the final loss function is described as:
%where $L_{ce}$ denotes the cross entropy criteria used for regularization and $L_{om}$ denotes the corresponding metric we explored. The hyper-parameter $\beta$ is set to 0.8 . \par%
In order to speed up the training process, we first train the network only with cross entropy (CE) criterion until it converges. Then we deploy our proposed MSCE criterion, using the CE criterion at the same time for regularization to reduce the overfiting tendency caused by sequence level training. The final loss function is described as
\vspace{-0.5em}
\begin{equation}
	L = \beta L_{\text{MSCE}} + (1-\beta) L_{\text{CE}}
\end{equation}
where $L_{\text{MSCE}}$ denotes our proposed MSCE criterion loss function and $L_{\text{CE}}$ denotes the CE loss function used for regularization. Usually, we scale the MSCE loss function by a user-specified constant $\beta$ = 0.8 during training.

\section{Experiments}
\vspace{-0.5em}
In this section, we introduce our corpus first, and then describe the experimental setup and results of our proposed methods. To the best of our knowledge, there does not exist any open-sourced corpus which is designed for SCR task particularly with confusing commands under practical scenarios. Note that the Google speech commands corpus~\cite{warden2018speech} is an open dataset, but its commands are quite simple and there is no obvious pronunciation confusion among the commands. Hence in this study, we design a reasonable dataset for experimentation.

\subsection{Corpus Details}
\vspace{-0.5em}
The corpus we used includes 41 Mandarin commands with different length (2-6 characters) for air conditioner control, including confusing commands with partial overlap in pronunciation. The training set is recorded on smart phones in near-field quiet environment, containing approximately 95k speech utterances from 1744 different speakers, in which each command has 1800-2500 utterances. We use pyroomacoustics~\cite{scheibler2018pyroomacoustics} to generate Room Impulse Response~(RIR) for data augmentation, augmenting the original training set 10 times to 950k utterances by adding noise of various kinds. Besides, another 400k non-command utterances are added into the training set as negative samples, which come from diverse human voice (such as movies, lectures and news) and audio of environment noise. For the development set, we randomly shuffle 50k utterances from the training set to construct it.

The evaluation set is recorded at different distances in several real rooms, with approximately 10k utterances from 55 speakers. Reading data are used as negative samples for computing False Alarm Rate~(FAR), which are randomly shuffled from the AISHELL-2 corpus~\cite{du2018aishell}.

\subsection{Experimental Setups}
\vspace{-0.5em}
All the experiments are conducted on the acoustic model with the same structure described in Section 2. The model includes 16 TDNN layer blocks, kernel size of each TDNN layer is 3, with 128 kernels. Dilation rate of each TDNN layer is [1, 2, 4, 4, 2, 1, 1, 2, 4, 4, 2, 1, 1, 2, 4, 4], respectively.
The 40-dimensional mel-filterbanks are used as input features. The same search graph is also used for all the experiments, and we dynamically adjust the pruning thresholds to draw receiver operating characteristic (ROC) curves. ROC curves and confusion errors at specific FAR, which we defined as the number of mis-classification samples whose decoding results are not negative, are reported in this paper. In order to ensure the reliability of the results and avoid random effects, all results are completed on the basis of model averaging.

\subsection{Comparison on Different Criteria}
\begin{figure}[!h]
	\vspace{-5pt}
	\centering 
	\includegraphics[width=1.0\linewidth]{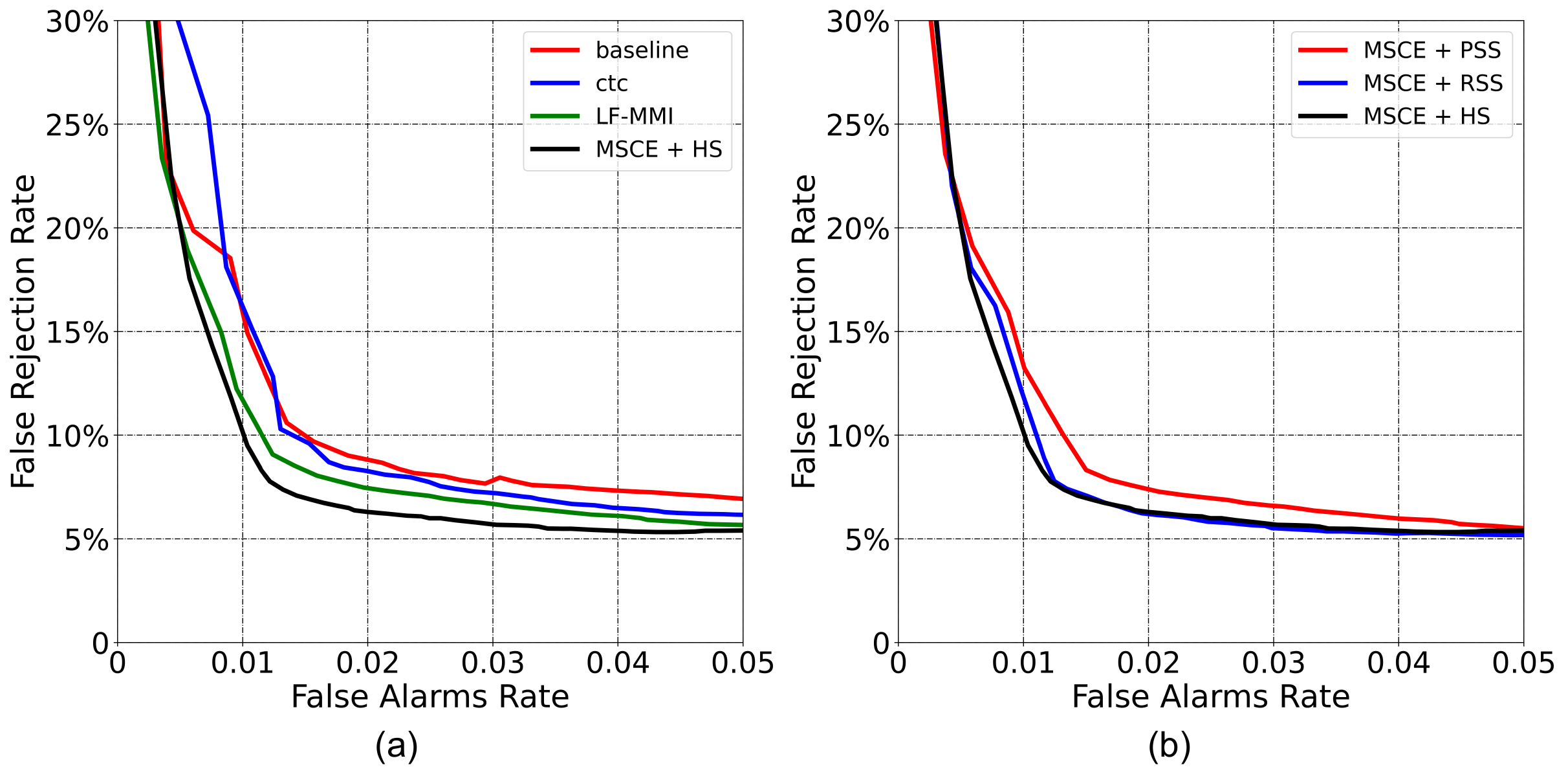}\vspace{-10pt}
	\caption{(a) shows the ROC curves Comparison of different criteria and (b) shows the ROC curves Comparison of different strategies using in MSCE criteria.}\vspace{-15pt}
	\label{fig2}
\end{figure}

\vspace{-0.5em}
We first show the performance of different criteria, including CE, CTC, lattice-free MMI~(LF-MMI) and our proposed MSCE criterion with HS strategy on ROC curves. The ROC curves are shown in Fig.~\ref{fig2}(a). Compared with the baseline CE criterion, our proposed MSCE criterion is superior with 33.7\% relative reduction at FRR at 0.01 FAR. Compared with the CTC criterion for ablation experiment, our proposed MSCE criterion is also superior. And we note that single CTC criterion does not always work well; it even hurts the model when the pruning threshold is relative tight at FAR under 0.01. We found that most of the errors are reject errors and it happen more for the evaluation utterances with low signal-to-noise ratio (SNR) in this case, because the peak posterior generated by CTC makes it prune the correct token prematurely. Compared with LF-MMI, a commonly used discriminative training criterion in speech recognition task and successfully used in keyword spotting as well~\cite{wang2020wake}, our proposed MSCE criterion is also superior with 12.8\% relative reduction at FRR at 0.01 FAR.

% \begin{table}[]
% 	\caption{FRR(\%) at different FAR for criteria under comparison. }\vspace{5pt}
% 	\label{tab:1}
% 	\centering
% 	\scalebox{0.9}{
% 		\renewcommand{\arraystretch}{1.3}
% 		\begin{tabular}{c|ccc|ccc}
% 			\toprule
% \multicolumn{1}{l|}{\multirow{2}{*}{FAR(\%)}} & \multirow{2}{*}{CE} & \multicolumn{1}{l}{\multirow{2}{*}{CTC}} & \multicolumn{1}{l|}{\multirow{2}{*}{LF-MMI}} & \multicolumn{3}{c}{MSCE} \\
% \multicolumn{1}{l|}{}                         &                     & \multicolumn{1}{l}{}                     & \multicolumn{1}{l|}{}                        & PSS     & RSS    & HS     \\ \hline
% 			1.0                    & 14.55    & 19.26   & 10.54    & 11.03    & 11.98    & \textbf{10.08}       \\ \hline
% 			2.0                    & 7.93     & 7.50    & 6.54     & 7.10     & 6.51     & \textbf{6.36}        \\ \hline
% 			5.0                    & 6.61     & 5.87    & 5.28     & 5.36     & 5.25     & \textbf{5.22}       \\ \hline
% 			Gain(\%)               & -        & -5.25   & 21.73    & 17.85    & 18.71    & \textbf{23.84}       \\ \bottomrule
% 	\end{tabular}}\vspace{-10pt}
% \end{table}

\begin{table}[]
	\caption{Relative improvement (\%) on confusion errors for different criteria under comparison.}%\vspace{5pt}
	\label{tab:3}
	\centering
	\scalebox{0.9}{
	\setlength{\tabcolsep}{2.5mm}{
		\renewcommand{\arraystretch}{1.3}
		\begin{tabular}{c|ccc|ccc}
			\toprule
\multicolumn{1}{l|}{\multirow{2}{*}{FAR}} & \multirow{2}{*}{CE} & \multicolumn{1}{l}{\multirow{2}{*}{CTC}} & \multicolumn{1}{l|}{\multirow{2}{*}{LF-MMI}} & \multicolumn{3}{c}{MSCE} \\
\multicolumn{1}{l|}{}                         &                     & \multicolumn{1}{l}{}                     & \multicolumn{1}{l|}{}                        & PSS     & RSS    & HS     \\ \hline
			0.01                    & -    & -8.65   & 5.12    & 10.7    & 12.01    & \textbf{12.17}       \\ \hline
			0.02                    & -     & 7.32    & 9.5     & 6.3     & \textbf{15.15}     & 13.13        \\ \hline
			0.05                    & -    & 19.49    & 17.81     & 25.57     & 27.25     & \textbf{29.55}       \\ \hline
			Gain              & -        & 6.05   & 10.81    & 14.19    & 18.13    & \textbf{18.28}       \\ \bottomrule
	\end{tabular}}}%\vspace{-10pt}
\end{table}

%\vspace{-10pt}

\subsection{Discussion on MSCE Training}
\vspace{-0.5em}
We also explore the strategies on confusing command selection for the denominator part of the MSCE criterion described in Section 3.3.
For all different strategies, we set the hyper-parameter $N=4$. As shown in Fig.~\ref{fig2}(b), HS is superior to RSS while RSS is superior to PSS. The result indicates that choosing commands from a whole command pool is essential as one of the sampling strategies, as it provides the proper class coverage, more in line with the definition of non-target command set in the MSCE criterion. On this basis, adding similar commands appropriately in the confusing command set can further improve the performance. 

% \begin{figure}[!h]
% 	\vspace{-5pt}
% 	\centering 
% 	\includegraphics[width=0.9\linewidth]{MSCE_2.pdf}\vspace{-10pt}
% 	\caption{ROC curves Comparison of different strategies using in MSCE criteria.}\vspace{-5pt}
% 	\label{fig3}
% \end{figure}

\subsection{Analysis on Confusion}
\vspace{-0.5em}
Another important purpose of our proposed MSCE criterion is to alleviate the confusion error in SCR. So we particularly evaluate the relative reduction on confusion error compared to other training criteria on the evaluation set at several important operate points. The results are shown in Table~\ref{tab:3}. We can see that our proposed MSCE criterion achieves the best performance while reducing 18.28\% confusion errors relatively. Meanwhile, CTC and LF-MMI can also get smaller gain of about 6\% and 10\% respectively.
Again, HS performs best for confusion reduction among different selection strategies for choosing confusing commands. RSS and HS perform almost the same with a gain of about 18\%, both of which are superior to PSS with gain of 14.19\%. This means that a proper class coverage is of great impact. And HS is slightly superior to RSS with the gain of 18.28\%, which proves that confusing commands with similar pronunciation can obtain a further gain on this basis.

\section{Conclusions}
\vspace{-0.5em}
In this paper, we propose a novel MSCE training criterion for SCR, aiming to improve command recognition performance as well as alleviating confusion errors among different commands without additional model parameters and time consumption. The MSCE criterion adopts CTC as sequence-level probability function and creates a confusing sequence set for each command in order to suppress the likelihood of non-target sequences during training. We evaluate our method on a sizable dataset, showing 18.28\% reduction in confusion errors and 33.7\% reduction in FRR at 0.01 FAR over the CE training baseline. Our proposed MSCE criterion is also superior to LF-MMI -- another popular discriminative training criterion.

\bibliographystyle{IEEEtran}
\bibliography{mybib}

% Generated by IEEEtran.bst, version: 1.13 (2008/09/30)
\begin{thebibliography}{10}
\providecommand{\url}[1]{#1}
\csname url@samestyle\endcsname
\providecommand{\newblock}{\relax}
\providecommand{\bibinfo}[2]{#2}
\providecommand{\BIBentrySTDinterwordspacing}{\spaceskip=0pt\relax}
\providecommand{\BIBentryALTinterwordstretchfactor}{4}
\providecommand{\BIBentryALTinterwordspacing}{\spaceskip=\fontdimen2\font plus
\BIBentryALTinterwordstretchfactor\fontdimen3\font minus
  \fontdimen4\font\relax}
\providecommand{\BIBforeignlanguage}[2]{{%
\expandafter\ifx\csname l@#1\endcsname\relax
\typeout{** WARNING: IEEEtran.bst: No hyphenation pattern has been}%
\typeout{** loaded for the language `#1'. Using the pattern for}%
\typeout{** the default language instead.}%
\else
\language=\csname l@#1\endcsname
\fi
#2}}
\providecommand{\BIBdecl}{\relax}
\BIBdecl

\bibitem{1989Continuous}
J.~R. Rohlicek, W.~Russell, S.~Roukos, and H.~Gish, ``Continuous hidden
  {M}arkov modeling for speaker-independent word spotting,'' in \emph{IEEE
  International Conference on Acoustics, Speech, and Signal Processing
  (ICASSP)}, 1989, pp. 627--630.

\bibitem{1990A}
R.~C. Rose and D.~B. Paul, ``A hidden {M}arkov model based keyword recognition
  system,'' in \emph{IEEE International Conference on Acoustics, Speech, and
  Signal Processing (ICASSP)}, 1990, pp. 129--132.

\bibitem{2016Multi}
S.~Panchapagesan, M.~Sun, A.~Khare, S.~Matsoukas, A.~Mandal, B.~Hoffmeister,
  and S.~Vitaladevuni, ``Multi-task learning and weighted cross-entropy for
  {DNN}-based keyword spotting.'' in \emph{ISCA Conference of the International
  Speech Communication Association (Interspeech)}, 2016, pp. 760--764.

\bibitem{2017Compressed}
M.~Sun, D.~Snyder, Y.~Gao, V.~K. Nagaraja, M.~Rodehorst, S.~Panchapagesan,
  N.~Strom, S.~Matsoukas, and S.~Vitaladevuni, ``Compressed time delay neural
  network for small-footprint keyword spotting.'' in \emph{ISCA Conference of
  the International Speech Communication Association (Interspeech)}, 2017, pp.
  3607--3611.

\bibitem{2018Monophone}
M.~Wu, S.~Panchapagesan, M.~Sun, J.~Gu, R.~Thomas, S.~N.~P. Vitaladevuni,
  B.~Hoffmeister, and A.~Mandal, ``Monophone-based background modeling for
  two-stage on-device wake word detection,'' in \emph{IEEE International
  Conference on Acoustics, Speech and Signal Processing (ICASSP)}, 2018, pp.
  5494--5498.

\bibitem{2014Small}
G.~Chen, C.~Parada, and G.~Heigold, ``Small-footprint keyword spotting using
  deep neural networks,'' in \emph{IEEE International Conference on Acoustics,
  Speech and Signal Processing (ICASSP)}, 2014, pp. 4087--4091.

\bibitem{0Convolutional}
T.~Sainath and C.~Parada, ``Convolutional neural networks for small-footprint
  keyword spotting,'' 2015.

\bibitem{wei2021end}
B.~Wei, M.~Yang, T.~Zhang, X.~Tang, X.~Huang, K.~Kim, J.~Lee, K.~Cho, and S.-U.
  Park, ``End-to-end transformer-based open-vocabulary keyword spotting with
  location-guided local attention,'' in \emph{ISCA Conference of the
  International Speech Communication Association (Interspeech)}, 2021, pp.
  361--365.

\bibitem{2006Connectionist}
A.~Graves, S.~Fern{\'a}ndez, F.~Gomez, and J.~Schmidhuber, ``Connectionist
  temporal classification: {L}abelling unsegmented sequence data with recurrent
  neural networks,'' in \emph{ACM International Conference on Machine Learning
  (ICML)}, 2006, pp. 369--376.

\bibitem{yan2020crnn}
H.~Yan, Q.~He, and W.~Xie, ``{CRNN-CTC} based {M}andarin keywords spotting,''
  in \emph{IEEE International Conference on Acoustics, Speech and Signal
  Processing (ICASSP)}, 2020, pp. 7489--7493.

\bibitem{2018Efficient}
S.~Myer and V.~S. Tomar, ``Efficient keyword spotting using time delay neural
  networks,'' in \emph{ISCA Conference of the International Speech
  Communication Association (Interspeech)}, 2018.

\bibitem{coucke2019efficient}
A.~Coucke, M.~Chlieh, T.~Gisselbrecht, D.~Leroy, M.~Poumeyrol, and T.~Lavril,
  ``Efficient keyword spotting using dilated convolutions and gating,'' in
  \emph{IEEE International Conference on Acoustics, Speech and Signal
  Processing (ICASSP)}, 2019, pp. 6351--6355.

\bibitem{lengerich2016endtoend}
C.~Lengerich and A.~Hannun, ``An end-to-end architecture for keyword spotting
  and voice activity detection,'' 2016.

\bibitem{wang2017smallfootprint}
Z.~Wang, X.~Li, and J.~Zhou, ``Small-footprint keyword spotting using deep
  neural network and connectionist temporal classifier,'' 2017.

\bibitem{2016Purely}
D.~Povey, V.~Peddinti, D.~Galvez, P.~Ghahremani, V.~Manohar, X.~Na, Y.~Wang,
  and S.~Khudanpur, ``Purely sequence-trained neural networks for asr based on
  lattice-free {MMI},'' in \emph{ISCA Conference of the International Speech
  Communication Association (Interspeech)}, 2016, pp. 2751--2755.

\bibitem{Chen_2018}
Z.~Chen, Y.~Qian, and K.~Yu, ``Sequence discriminative training for deep
  learning based acoustic keyword spotting,'' \emph{Elsevier Speech
  Communication}, vol. 102, pp. 100--111, 2018.

\bibitem{2002Minimum}
D.~Povey and P.~C. Woodland, ``Minimum phone error and {I}-smoothing for
  improved discriminative training,'' in \emph{IEEE International Conference on
  Acoustics, Speech, and Signal Processing (ICASSP)}, vol.~1, 2002, pp. I--105.

\bibitem{2016Minimum}
T.~Hori, C.~Hori, S.~Watanabe, and J.~R. Hershey, ``Minimum word error training
  of long short-term memory recurrent neural network language models for speech
  recognition,'' in \emph{IEEE International Conference on Acoustics, Speech
  and Signal Processing (ICASSP)}, 2016, pp. 5990--5994.

\bibitem{guo2020efficient}
J.~Guo, G.~Tiwari, J.~Droppo, M.~Van~Segbroeck, C.-W. Huang, A.~Stolcke, and
  R.~Maas, ``Efficient minimum word error rate training of {RNN-T}ransducer for
  end-to-end speech recognition,'' in \emph{ISCA Conference of the
  International Speech Communication Association (Interspeech)}, 2020.

\bibitem{0Investigating}
S.-H. Liu, F.-H. Chu, S.-H. Lin, and B.~Chen, ``Investigating data selection
  for minimum phone error training of acoustic models,'' in \emph{IEEE
  International Conference on Multimedia and Expo (ICME)}, 2007, pp. 348--351.

\bibitem{1992Discriminative}
B.-H. Juang and S.~Katagiri, ``Discriminative learning for minimum error
  classification (pattern recognition),'' \emph{IEEE Transactions on Signal
  Processing (TSP)}, vol.~40, no.~12, pp. 3043--3054, 1992.

\bibitem{4032780}
E.~McDermott, T.~J. Hazen, J.~Le~Roux, A.~Nakamura, and S.~Katagiri,
  ``Discriminative training for large-vocabulary speech recognition using
  minimum classification error,'' \emph{IEEE Transactions on Audio, Speech, and
  Language Processing (TASLP)}, vol.~15, no.~1, pp. 203--223, 2006.

\bibitem{wang2020wake}
Y.~Wang, H.~Lv, D.~Povey, L.~Xie, and S.~Khudanpur, ``Wake word detection with
  alignment-free lattice-free {MMI},'' in \emph{ISCA Conference of the
  International Speech Communication Association (Interspeech)}, 2020.

\bibitem{du2018aishell}
J.~Du, X.~Na, X.~Liu, and H.~Bu, ``Aishell-2: {T}ransforming {M}andarin {ASR}
  research into industrial scale,'' \emph{arXiv preprint arXiv:1808.10583},
  2018.

\bibitem{scheibler2018pyroomacoustics}
R.~Scheibler, E.~Bezzam, and I.~Dokmani{\'c}, ``Pyroomacoustics: {A} python
  package for audio room simulation and array processing algorithms,'' in
  \emph{IEEE International Conference on Acoustics, Speech and Signal
  Processing (ICASSP)}, 2018, pp. 351--355.

\bibitem{warden2018speech}
P.~Warden, ``Speech commands: {A} dataset for limited-vocabulary speech
  recognition,'' \emph{arXiv preprint arXiv:1804.03209}, 2018.

\bibitem{vincent2008extracting}
P.~Vincent, H.~Larochelle, Y.~Bengio, and P.-A. Manzagol, ``Extracting and
  composing robust features with denoising autoencoders,'' in \emph{ACM
  International Conference on Machine Learning (ICML)}, 2008, pp. 1096--1103.

\end{thebibliography}

% \begin{thebibliography}{9}
% \bibitem[1]{Davis80-COP}
%   S.\ B.\ Davis and P.\ Mermelstein,
%   ``Comparison of parametric representation for monosyllabic word recognition in continuously spoken sentences,''
%   \textit{IEEE Transactions on Acoustics, Speech and Signal Processing}, vol.~28, no.~4, pp.~357--366, 1980.
% \bibitem[2]{Rabiner89-ATO}
%   L.\ R.\ Rabiner,
%   ``A tutorial on hidden Markov models and selected applications in speech recognition,''
%   \textit{Proceedings of the IEEE}, vol.~77, no.~2, pp.~257-286, 1989.
% \bibitem[3]{Hastie09-TEO}
%   T.\ Hastie, R.\ Tibshirani, and J.\ Friedman,
%   \textit{The Elements of Statistical Learning -- Data Mining, Inference, and Prediction}.
%   New York: Springer, 2009.
% \bibitem[4]{YourName17-XXX}
%   F.\ Lastname1, F.\ Lastname2, and F.\ Lastname3,
%   ``Title of your INTERSPEECH 2022 publication,''
%   in \textit{Interspeech 2022 -- 23\textsuperscript{rd} Annual Conference of the International Speech Communication Association, September 18-22, Incheon, Korea, Proceedings, Proceedings}, 2022, pp.~100--104.
% \end{thebibliography}

\end{document}